\documentclass[10pt, conference]{IEEEtran} 
\usepackage{graphicx}
\usepackage{color}
\usepackage{amssymb}
\usepackage{epsfig}
\begin{document}
\bibliographystyle{IEEEtran} 

\title{On the Capacity Bounds of Undirected Networks}  
\author{Ali Al-Bashabsheh and Abbas Yongacoglu           
              \\School of Information Technology and Engineering
              \\University of Ottawa, Ottawa, Canada
              \\\{aalba059, yongacog\}@site.uottawa.ca
}

\date{}
\maketitle

\begin{abstract}
 In this work we improve on the bounds presented in \cite{Li2004:1} for network coding gain in the undirected case. A tightened bound for the undirected multicast problem with three terminals is derived. An interesting result shows that with fractional routing, routing throughput can achieve at least 75\% of the coding throughput. A tighter bound for the general multicast problem with any number of terminals shows that coding gain is strictly less than 2. Our derived bound depends on the number of terminals in the multicast network and approaches 2 for arbitrarily large number of terminals. 

\end{abstract}



\section{Introduction}

Network coding gain, $\mathfrak{G}$, is defined as the improvement in network throughput due to coding compared to routing throughput. It is well known that network coding increases the space over which throughput is maximized and thus provides a capacity at least as high as the routing one. In this work, capacity of a network refers to the maximum achievable throughput under a certain coding scheme (routing can be considered as a coding scheme with identity mapping). For a set of sinks in a directed multicast network, it was shown in \cite{Yeung2000:1} that if the network can achieve a certain throughput to each receiver individually, then it can achieve the same throughput to all the sinks simultaneously by allowing coding at intermediate nodes and thus, achieve a throughput gain. This has ignited an area of research trying to answer many questions; one of them is how much gain is possible? There has been some instances in the literature where the network coding gain can be unbounded for directed networks \cite{Sanders2005:1} \cite{Fragouli2006:1}. Relating coding gain to the integrality gap of linear programming formulation for minimum weight Steiner tree \cite{Agarwal2004:1}, further examples of directed networks with arbitrarily high coding gain can be obtained \cite{Zosin2002:1} \cite{Halperin2003:1}. For undirected networks, it was shown in \cite{Li2004:1} that network coding gain is bounded by $2$ when half integer routing is possible. Such a dramatic difference between directed and undirected multicast networks results because directed networks can be oriented in a way such that routing can offer little compared to coding. In this work we further investigate achievable routing throughputs and derive tighter bounds on the possible coding gain. 

This work is organized as follows. Section II introduces some necessary definitions. In Section III we use a Steiner tree packing argue to present bounds on the achievable routing throughput and derive upper bounds on coding gain. 

\section{Definitions}

An undirected network $G$ on $V$ nodes and $E$ links can be modeled as an undirected graph $G(V,E)$ that might contain multiple edges, i.e. a multigraph. At certain places we might refer to the sets of vertices and edges of a graph $H$ (especially if they are not explicitly specified) as $V(H)$ and $E(H)$, respectively. A multicast network with a source node $s \in V$ and a set of sinks ${\cal R} = \{R_{1}, \ldots, R_{|{\cal R}|}\} \subseteq V-s $ is a network where $s$ broadcasts data to all the receivers in $\cal R$. At certain places in this work we might not distinguish between a source or a sink node and simply refer to such a node as a \textit{terminal}. If ${|\cal R|} = V - 1$ 
then we call $G$ a \textit{broadcast} 
network.

We assume that edges capacities are defined over the same base as the source symbols. In other words, an edge with capacity $C$ can carry at most $C$ symbols. Integer routing throughput refers to the routing throughput achieved by routing scalar symbols through edges of the network. Considering multiple time units in a network, fractional routing throughput denotes the throughput achieved by upscaling edge capacities by $n$ divided by $n$. Such a routing scheme may also be referred to as vector routing and it was shown in \cite{Riis2003:1} that doubling edges capacities can result in a throughput more than twice as good. If we let $h$ denote the number of symbols a multicast network can deliver from the source to all sinks, we can define coding and fractional and integer routing capacities as: 
\begin{eqnarray}
&\gamma& \!\!\! = \mbox{sup} \left\{ {h}/{n} \in {\mathbb{Q^{+}}}: {h}/{n} \mbox{ is an achievable coding rate} \right\} \nonumber \\ 
&\pi&\!\!\!\!\!\!\!_{f} = \mbox{sup} \left\{ h/n \in {\mathbb{Q^{+}}}: h/n \mbox{ is an achievable routing rate} \right\} \nonumber \\
&\pi&\!\!\!\!\!\!\!_{i} = \mbox{sup} \left\{ h \in {\mathbb{Z^{+}}}: h \mbox{ is an achievable integer routing rate} \right\} \nonumber 
\end{eqnarray}
where ${\mathbb{Z^{+}}}$ and ${\mathbb{Q^{+}}}$ are the sets of positive integers and positive rational numbers, respectively. A fractional routing scheme with $n=2$ will be referred to as half integer routing and the corresponding capacity will be denoted by $\pi_{\frac{1}{2}}$. In the following, if a statement is true for both integer and fractional routing, we may drop the subscripts and simply write $\pi$. The gain in throughput due to coding compared to integer, half-integer and fractional routing is defined as $\mathfrak{G}_{i} = \frac{\gamma}{\pi_{i}}$, $\mathfrak{G}_{\frac{1}{2}} = \frac{\gamma}{\pi_{\frac{1}{2}}}$ and $\mathfrak{G}_{f} = \frac{\gamma}{\pi_{f}}$, respectively.

For a graph $G = (V,E)$, an induced subgraph $T$ is called a \textit{spanning tree} if and only if $T$ is a tree and $V(T) = V(G)$. For $A \subseteq V$, a subtree $T$ of $G$ is called an \textit{$A$-Steiner tree} ($A$-Spanning tree) if and only if $A \subseteq V(T)$. Steiner tree packing refers to finding the maximum number of edge disjoint $A$-Steiner trees in $G$.

\textit{Edge connectivity}, $\lambda(G)$, of a connected graph $G$ is the size of the smallest set $F \subseteq E(G)$ such that $G - F$ is disconnected ($G - F$ denotes the graph induced from $G$ by deleting all edges $e \in F$). A graph $G$ is $l$-edge connected for any integer $l \leq \lambda(G)$.
 For $A \subseteq V(G)$, $\lambda_{G}(A)$ denotes the edge connectivity of $A$ in $G$. The minimum size of a cut between a pair of vertices $u, v \in V(G)$ is the edge connectivity, $\lambda_{G}(u,v)$, of $u$ and $v$ in $G$ which by Menger's theorem equals the maximum number of edge disjoint paths between $u$ and $v$. A \textit{cut-edge} (bridge) is an edge $e \in E(G)$ such that $G-e$ is disconnected (clearly, $\lambda(G)$ must be 1 for such an edge to exist). The degree of $u \in V$ is the number of incident edges to $u$ and is denoted as $d(u)$. If the underlying graph $G$ is clear from the context, we might drop the subscript ${G}$ from the previous quantities.

Let $e = rx$ and $f = xt$ be two edges in $G(V,E)$. \textit{Splitting off} the pair of edges $e$ and $f$ refers to deleting $e$ and $f$ and introducing a new edge $g = rt$. We refer to the pair $e,f$ and $g$ as the splitted and the splitting edges, respectively. We also denote the resulting graph after splitting off edges $e,f$ as $G^{ef}$. A pair of edges $e,f$ incident with $x$ is \textit{admissible} if $\lambda_{G^{ef}}(u,v) = \lambda_{G}(u,v)$ for every $u \neq v \in V-x$. An incident edge to $x$ is admissible if it belongs to an admissible pair, otherwise it is non-admissible. A complete splitting at $x \in V$ (when $d(x)$ is even) refers to: 1) repeatedly splitting off pairs of incident edges to $x$ until $x$ is isolated, and 2) deleting $x$. We denote the graph resulting from a complete splitting at $x$ as $G^{x}$ and call such a splitting a \textit{suitable} one if every splitting in step 1) is admissible, i.e. all edges incident to $x$ belong to \textit{disjoint} admissible pairs.

\section{Bounds on Achievable Rates for Undirected Networks}
For network $G$ modeled as an undirected graph $G(V,E)$ and a set of terminals $A \subseteq V$, using a max-flow min-cut argument, the network coding capacity is trivially upper bounded by $\lambda(A)$. Note that unlike directed networks (where the min-cut throughput is always achievable via coding over a sufficiently large field \cite{Yeung2000:1} \cite{Yeung2003:1}) in the undirected case a throughput of $\lambda(A)$ might not be achievable (see example 2). It is also clear that $\pi \leq \gamma$ since routing is a special type of coding where only repetition and forwarding are allowed at relay nodes. If $\lambda(A) = 1$, then it is clear that $\gamma = \pi = 1$ and hence we can always assume $\lambda(A) \geq 2$. On the other hand, if $\lambda(G) = 1$ and $\lambda(A)  \geq 2$ then $V(G)$ can be partitioned into two disjoint subsets $V(G_{1})$ and $V(G_{2})$ with $A \subseteq V(G_{1})$ and $G_{1}$ and $G_{2}$ connected via a {cut-edge}. In this case we can always delete $G_{2}$ without affecting the throughput. In conclusion, we can always assume that $G$ is 2-edge connected, i.e. $G$ contains no cut-edges.

\subsection{Multicast Networks with Three Terminals}

Let $G = (V,E)$ be a graph representing a multicast network with a set of terminals $A = \{s,R_{1},R_{2}\} \subseteq V$, the following theorem provides a lower bound on the routing capacity, 
\\
\textbf{Theorem 1}: \textit{For an undirected multicast network with three terminals, the integer, half integer and fractional routing capacities are bounded as
\begin{eqnarray}
\pi_{i} & \geq & \left \lfloor \frac{6\lambda(A) - 3}{8} \right\rfloor \nonumber 
\end{eqnarray}
\begin{eqnarray}
\pi_{\frac{1}{2}} & \geq & \frac{1}{2} \left \lfloor \frac{12\lambda(A) - 3}{8} \right\rfloor \nonumber  \\
\pi_{f} & \geq & \frac{3 \lambda(A)}{4} - \epsilon \nonumber
\end{eqnarray}
where $\epsilon$ is arbitrarily small for arbitrarily large $n$ 
and $\lambda(A)$ is the connectivity of $A$ in $G$}.
\\
\textbf{Proof}: The proof makes use of the following theorem \\
\textbf{Theorem 2} (Kriesell, \cite{Kriesell2003:1}).
\textit{ For any integer $k \geq 1$, let $A = \{v_{1}, v_{2}, v_{3}\}$ be $ \left\lfloor \frac{8k+3}{6} \right\rfloor $-edge connected in $G$, then there exists a system of $k$ edge disjoint $A$-Steiner trees in $G$}.
\\
Let $A = \{v_{1}, v_{2}, v_{3}\}$ be the set of terminals and let $A$ be $\lambda(A)$-edge connected in $G$. Noting that any positive integer, $a$, can be written as
\begin{eqnarray}
a = dq + b; \hspace{.2cm} \mbox{for some choice} \hspace{.2cm} q \in \{0\} \cup { \mathbb{Z^{+}} }, \nonumber \\ 
d \in \{1,2,\ldots, a \} \hspace{.2cm} \mbox{and} \hspace{.2cm} b \in \{0,1,2,\ldots, d-1 \}  
\label{eq:any_int}
\end{eqnarray}
Consider,
\begin{eqnarray}
\left\lfloor \frac{8k+3}{6} \right\rfloor = \left\lfloor \frac{(6+2)k+3}{6} \right\rfloor = k + \left\lfloor \frac{2k+3}{6} \right\rfloor,  \hspace{.2cm} \forall \hspace{.1cm}  k  \in { \mathbb{Z^{+}} }
\nonumber
\end{eqnarray}
From (\ref{eq:any_int}) with $d=3$, we can write $k=3q+m$, $m \in \{0,1,2\}$.
Thus,
\begin{eqnarray}
\left\lfloor \frac{8k+3}{6} \right\rfloor &=& 3q + m + \left\lfloor \frac{6q+2m + 3}{6} \right\rfloor   \cr
&=& 4q + m + \left\lfloor \frac{2m+3}{6} \right\rfloor, \hspace{.5cm} m \in \{0,1,2\} \cr
&=& \left\{ 
\begin{array}{lc}      4q, & m=0 \\ 4q+1, & m=1 \\     4q+3, & m=2 
\end{array} \right. \nonumber
\end{eqnarray}
Thus, if $\lambda(A)$ is any positive integer such that $\lambda(A) = 4q+j$, $j \in \{0,1,3\}$, we can write $\lambda(A) = \left\lfloor \frac{8k+3}{6} \right\rfloor$ for some positive integer $k$. The only class of positive integers left is the one such that $\lambda(A) = 4q+2$. For this choice of integers, we can write $\lambda(A) = 4q+1 +1 = \left\lfloor \frac{8k+3}{6} \right\rfloor + 1$ for some positive integer $k$. Hence, for any positive integer, $\lambda(A)$, there exist a positive integer $k$ for which $\lambda(A) = \left\lfloor \frac{8k+3}{6} \right\rfloor + \delta$, where $k$ is the \textit{largest} integer such that $\left\lfloor \frac{8k+3}{6} \right\rfloor \leq \lambda(A)$ and $\delta \in \{0,1\}$.

For $\delta = 0$, Theorem 2 ensures the existence of $k$ edge disjoint $A$-Steiner trees. Thus,
\begin{eqnarray}
\lambda(A) = \left\lfloor \frac{8k+3}{6} \right\rfloor \leq \frac{8k+3}{6} \nonumber
\end{eqnarray} 
Which leads to
\begin{eqnarray}
k \geq \frac{6\lambda(A)-3}{8} \geq \left\lfloor \frac{6\lambda(A)-3}{8} \right\rfloor
\label{eq:delta0_bound_k}
\end{eqnarray}
For $\delta = 1$, Theorem 2 also ensures the existence of at {least} $k$ edge disjoint $A$-Steiner trees, where
$\lambda(A) = \left\lfloor \frac{8k+3}{6} \right\rfloor + 1$
, from which we first obtain
\begin{eqnarray}
k < \frac{6\lambda(A)-3}{8}
\label{eq:delta1_upperbound_k}
\end{eqnarray}
we also obtain the following lower bound on $k$
\begin{eqnarray}
k \geq \frac{6(\lambda(A)-1)-3}{8}
\label{eq:delta1_lowerbound_k}
\end{eqnarray}
Upon combining (\ref{eq:delta1_upperbound_k}) and (\ref{eq:delta1_lowerbound_k}), $k$ can be bounded as,
\begin{eqnarray}
\frac{6\lambda(A)-3}{8} - \frac{6}{8} \leq k < \frac{6\lambda(A)-3}{8} \nonumber
\end{eqnarray}
Noting that
\begin{eqnarray}
\frac{6\lambda(A)-3}{8} \!=\! \left\lfloor \frac{6\lambda(A)-3}{8} \right\rfloor \! + \frac{\Delta}{8}, \hspace{.1cm} \Delta \! \in \! S \! \subset \{1, 2, \ldots, 7\}
\label{eq:Delta_3ter}
\end{eqnarray}
(More specifically, it can be shown that for any integer $\lambda(A)$, i.e. $\delta$ can be $0$ or $1$,  $\Delta \in \{1,3,5,7\}$, see part-A of the Appendix).
Thus the inequality for $k$ becomes
\begin{eqnarray}
\left\lfloor \frac{6\lambda(A)-3}{8} \right\rfloor - \frac{6-\Delta}{8} \leq k < \left\lfloor \frac{6\lambda(A)-3}{8} \right\rfloor + \frac{\Delta}{8}
\end{eqnarray}
Note that $k$ is an integer, hence the previous inequality indicates that $k = \left\lfloor \frac{6\lambda(A)-3}{8} \right\rfloor$ \textit{except} for $\Delta = 7$, where the inequality has no valid solution. Therefore, our task is to prove that $\delta = 1$ dismisses $\Delta = 7$. To prove this, note that for $\delta = 1$ we have $\lambda(A) = 4q + 2$, $q \in \{0\} \cup {\mathbb{Z^{+}}}$. Hence,
\begin{eqnarray}
\frac{6\lambda(A)-3}{8}  &=&  \frac{3\times8q+9}{8}  \cr
&=& 3q+1+  \frac{1}{8}  \Longrightarrow \Delta = 1  \nonumber
\end{eqnarray} 
Which shows that $\Delta$ is always 1 for $\delta = 1$. From this we conclude that the solution presented earlier is a valid one. Combining this result with the one obtained for the case of $\delta = 0$ in (\ref{eq:delta0_bound_k}), we can write
\begin{eqnarray}
k \geq \left\lfloor \frac{6\lambda(A)-3}{8} \right\rfloor, \hspace{1cm} \mbox{for any} \hspace{.2cm} \lambda(A) \in {\mathbb{Z^{+}}}
\end{eqnarray}
Sending one symbol on each tree, an integral routing throughput of $k$ is achievable. Since $\pi_{i}$ is the supreme of all achievable rates, thus $\pi_{i} \geq k$ and the first bound results. 
If half-integer routing is allowed, we can up scale the edge capacity by 2 and then divide the integer part of the result by 2. Since this is an achievable routing rate, it represents a lower bound on the half integer routing capacity and thus proves the second part of the theorem. For fractional routing, we upscale edge capacities by $n$ and then divide the integer part of the result by $n$, i.e. $\pi_{f} \geq \frac{3\lambda(A)}{4} - \frac{3+\Delta}{8n}$. For sufficiently large $n$, the second part approaches zero and the third bound results. 
$\blacksquare$ 
\\
\textbf{Corollary 1}: \textit{For a three terminal multicast network, the coding gain is bounded as}
\begin{eqnarray}
{\mathfrak{G}}_{i} \!\! & \leq & \!\! \frac{\lambda(A)}{\left\lfloor \frac{6\lambda(A)-3}{8} \right\rfloor} \nonumber \\ 
{\mathfrak{G}}_{\frac{1}{2}} \!\! & \leq & \!\! \frac{2\lambda(A)}{\left\lfloor \frac{12\lambda(A)-3}{8} \right\rfloor} \nonumber \\ 
{\mathfrak{G}}_{f} \!\! & \leq & \!\! \frac{4}{3} + \epsilon^{\prime} \nonumber
\end{eqnarray}
\textbf{Proof}: ${\mathfrak{G}} = \frac{\gamma}{\pi} \leq \frac{\lambda(A)}{\pi} \leq \frac{\lambda(A)}{\mbox{\small Lower bound on } \pi}$. Applying the appropriate lower bound from theorem 1 and the corollary results.

From corollary 1, it can be seen that the coding gain is always bounded. 
 More specifically, ${\mathfrak{G}}_{i}$ is bounded by $3$ for $\lambda(A) \geq 2$ while ${\mathfrak{G}}_{\frac{1}{2}}$ is strictly less than $2$ for $\lambda(A) >2$. The bound for ${\mathfrak{G}}_{f}$ indicates that with fractional routing, 75\% of the throughput achievable by coding is always achievable via routing.

\subsection{Multicast Networks with Arbitrary Number of Terminals}
Let $G$ be a multicast network modeled as an undirected graph $G(V,E)$ with a set of $N-1$ receivers ${\cal R} = \{R_{1}, \ldots, R_{N-1}\}$ and thus a set of terminals $A = s \cup {\cal R} \subseteq V$. The set $X = V - A$ represents the set of non-terminal (relay) nodes in $G$. 
\\
\textbf{Lemma 1}: \textit{If $G^{x}$, the graph obtained from $G$ by performing a suitable complete splitting at $x \in V-A$, contains a system of $k$ edge disjoint $A$-Steiner trees, then $G$ contains $k$ edge disjoint $A$-steiner trees}.
\\
\textbf{Proof}: Let $\cal T$ be a set of $k$ disjoint $A$-Steiner trees in $G^{x}$. If $d(x) = 2$, let $w$ be the added edge after splitting off edges $e$ and $f$. For any $T \in {\cal T}$, if $T$ contains $w$ then $T-w$, the edges $e,f$ and the vertex $x$ form an $A$-Steiner tree $T^{\prime}$ in $G$ and the set $({\cal T} - T) \cup T^{\prime}$ forms a set of $k$ edge disjoint $A$-Steiner trees in $G$. If no $T \in {\cal T}$ contains $w$, then $\cal T$ is a set of $A$-Steiner trees in $G$. If $d(x) \neq 2$, let $W$ be the set of splitting edges and $e(w),f(w)$ be the splitted pair by $w$. Also let ${\cal T}_{W} = \{T \in {\cal T}: T \cap W \neq \emptyset \}$. For every $T \in {\cal T}_{W}$, the tree $T^{\prime}$ formed by $T-W$, $\bigcup_{w \in W \cap E(T)}{g_{w}}$ and $x$ is an $A$-Steiner tree in $G$, where $g_{w} \in \{ \emptyset, \{e(w)\}, \{f(w)\}, \{e(w),f(w)\} \}$. Let $\cal T^{\prime}$ be the set of such trees, then $ ({\cal T - T}_{W}) \cup \cal T^{\prime} $ forms a set of $k$ edge disjoint $A$-Steiner trees in $G$.
\\
\textbf{Alternative proof}: Since splitting off does not increase connectivity and complete splitting is a series of splitting off's then $\lambda_{G^{x}}(u,v) \leq \lambda_{G}(u,v)$ for all distinct $u, v \in V(G) - x$, with equality if and only if the splitting is suitable. Thus, if $G^{x}$ contains $k$-edge disjoint $A$-Steiner trees, then $G$ does as well.

The following theorem provides a lower bound on the fractional routing capacity, $\pi_{f}$.
\\
\textbf{Theorem 3}: \textit{For a multicast network represented by an undirected graph $G(V,E)$ with a set of terminals $A \subseteq V$
\begin{eqnarray}
\pi_{f} \geq \frac{\lambda(A)}{2} \left( \frac{|A|}{|A|-1} \right) - \epsilon  \nonumber
\end{eqnarray}
where $\epsilon$ is arbitrarily small for arbitrarily large $n$ 
and $\lambda(A)$ is the connectivity of $A$ in $G$}. \\
\textbf{Proof}: The proof uses the following two theorems \\
\textbf{Theorem 4} (Kriesell, \cite{Kriesell2003:1}): \textit{For any $l,k \geq 2$, let $G(V,E)$ be an $\left\lfloor  2k \frac{l-1}{l} + \frac{l-2}{l} \right\rfloor$-edge connected graph on $l$ vertices. Then there exist a system of $k$ edge disjoint spanning trees in $G$}. \\
\textbf{Theorem 5} (Frank, \cite{Frank1992:1}): \textit{Let $G(V+x,E)$ be a connected graph on $V+x$, $d(x) \neq 3$ and no cut-edge is incident to $x$. Then there exist $\left\lfloor \frac{d(x)}{2} \right\rfloor$ disjoint admissible pairs at $x$}. 

Consider a complete splitting at every relay node $x \in V-A$. Since $G$ is $2$-edge connected, then there are no cut-edges in $G$. From theorem 5, if $d(x)$ is even then all incident edges to $x$ can be partitioned into disjoint admissible pairs. Since half integer routing is allowed, we can make $d(x)$ even for every $x \in V-A$ by upscaling the capacities of the edges of $G$ by 2 and then downscaling the solution by 2. Thus a suitable complete splitting exists at all relay nodes in $G$. Let $G^{\prime}$ be the graph obtained after such splitting, then $V(G^{\prime}) = A$ and an $A$-Steiner tree in $G^{\prime}$ is a spanning tree in $G^{\prime}$. 

Next we prove that $\lambda(A)$ being any integer, it can be written as
\begin{eqnarray}
\lambda(A) = \left\lfloor  \frac{2k (|A|-1) + |A|-2}{|A|} \right\rfloor + \delta
\label{eq:lambda_general1}
\end{eqnarray}
for any $|A| \geq 2$ and some integer $k$, where $\delta \in \{0,1\}$. Let $f_{|A|}(k) = \left\lfloor  2k \frac{|A|-1}{|A|} + \frac{|A|-2}{|A|} \right\rfloor$, it is easy to check that $f_{|A|}(k+1) - f_{|A|}(k) = 2 + \left\lfloor \frac{\Delta - 2}{|A|} \right\rfloor$, $\Delta \in \{0,1, \ldots, |A|-1\}$. Noting that $\left\lfloor \frac{\Delta - 2}{|A|} \right\rfloor \in \{-1,0\}$ then, $f_{|A|}(k+1) - f_{|A|}(k) \in \{1,2\}$. If there exist an integer $k$ such that $\lambda(A) = f_{|A|}(k)$ then $\delta = 0$ and we are done. Otherwise, we choose $k$ as the largest integer such that $\lambda(A) > f_{|A|}(k)$. By the choice of $k$, $\lambda(A) < f_{|A|}(k+1)$ and since $f_{|A|}(k+1) - f_{|A|}(k) \leq 2$ for any $|A|$ then $\lambda(A) = f_{|A|}(k) + 1$. Thus, for any integer $\lambda(A)$, we can write $\lambda(A) = f_{|A|}(k) + \delta$, $\delta \in \{0,1\}$, proving the claim in(\ref{eq:lambda_general1}).

For $\lambda(A) = \left\lfloor \frac{2k(|A|-1) + |A|-2}{|A|} \right\rfloor + \delta$, theorem 4 ensures the existence of $k$ edge disjoint spanning trees in $G^{\prime}$, where for $\delta = 0$ we have $k \geq \left\lfloor \frac{|A|\lambda(A) - |A|+2}{2(|A|-1)} \right\rfloor$. If $\delta = 1$, then $\lambda(A) = \left\lfloor \frac{2k(|A|-1)+|A|-2}{|A|}\right\rfloor +1 > \frac{2k(|A|-1)+|A|-2}{|A|}$ which results in $k < \left\lfloor \frac{|A|\lambda(A) - |A|+2}{2(|A|-1)} \right\rfloor + \frac{\Delta^{\prime}}{2(|A|-1)}$, $\Delta^{\prime} \in \{0,1,\ldots, 2(|A|-1)-1\}$. 
Also $\lambda(A)-1 = \left\lfloor \frac{2k(|A|-1) + |A|-2}{|A|} \right\rfloor \leq \frac{2k(|A|-1) + |A|-2}{|A|}$ results in $ k \geq \left\lfloor \frac{|A|\lambda(A) - |A|+2}{2(|A|-1)} \right\rfloor + \frac{\Delta^{\prime}-|A|}{2(|A|-1)} $.
 Upon combining the above two limits of $k$, we can write $  \left\lfloor \frac{|A|\lambda(A) - |A|+2}{2(|A|-1)} \right\rfloor + \frac{\Delta^{\prime}-|A|}{2(|A|-1)} \leq   k < \left\lfloor \frac{|A|\lambda(A) - |A|+2}{2(|A|-1)} \right\rfloor + \frac{\Delta^{\prime}}{2(|A|-1)}$ which indicates that a valid solution for $k$ exists as long as $\Delta^{\prime} \in \{1,2,\ldots,|A|\}$. But $\delta = 1$ implies that that this is the case (part-B of the Appendix). Thus for $\delta = 1$, $k = \left\lfloor \frac{|A|\lambda(A) - |A|+2}{2(|A|-1)} \right\rfloor$. Combining this with the result obtained for $\delta = 0$, we conclude that for any integer $\lambda(A) \geq 2$, the number of edge-disjoint $A$-Steiner trees in $G^{\prime}$ is $k \geq \left\lfloor \frac{|A|\lambda(A) - |A|+2}{2(|A|-1)} \right\rfloor$.

By Lemma 1, if $G^{\prime}$ contains $k$ edge disjoint $A$-Steiner trees then $G$ does, provided that a suitable complete splitting exists at every relay node $x \in V-A$. As it was shown at the beginning of this argument, such a splitting exist if we upscale the capacity of every edge in $G$ by 2 and then downscale our result by 2. Thus, we conclude that $G$ contains $k$ edge disjoint $A$-Steiner trees, where $k \geq \frac{1}{2} \left\lfloor \frac{2|A|\lambda(A) - |A|+2}{2(|A|-1)} \right\rfloor $. Since this represents an achievable throughput, it also represents a lower bound on the half integer routing capacity. For fractional routing, we upscale the edge capacities by $n$ and divide the result by $n$ and for sufficiently large $n$ we obtain $\pi_{f} \geq \frac{\lambda(A)}{2} \left( \frac{|A|}{|A|-1} \right) - \epsilon$. $\blacksquare$ 
\\
\textbf{Corollary 2}: \textit{For a multicast network represented by an undirected graph $G(V,E)$ with a set of terminals $A \subseteq V$}
\begin{eqnarray}
{\mathfrak G}_{f} \leq 2 \left( \frac{|A|-1}{|A|} \right) + \epsilon^{\prime} \nonumber
\end{eqnarray}
\textbf{Proof}: Proof is similar to the one of corollary 1.
\subsection{Examples}
\begin{figure}[htbp]
		\centering \setlength{\unitlength}{0.38cm}
		\begin{picture}(18,11)(1,0)
		\put(4,10.5){$s$} \put(15.8,10.5){$s$}
		\put(-3,0){\includegraphics[width = 5.7cm,height = 4.56cm]{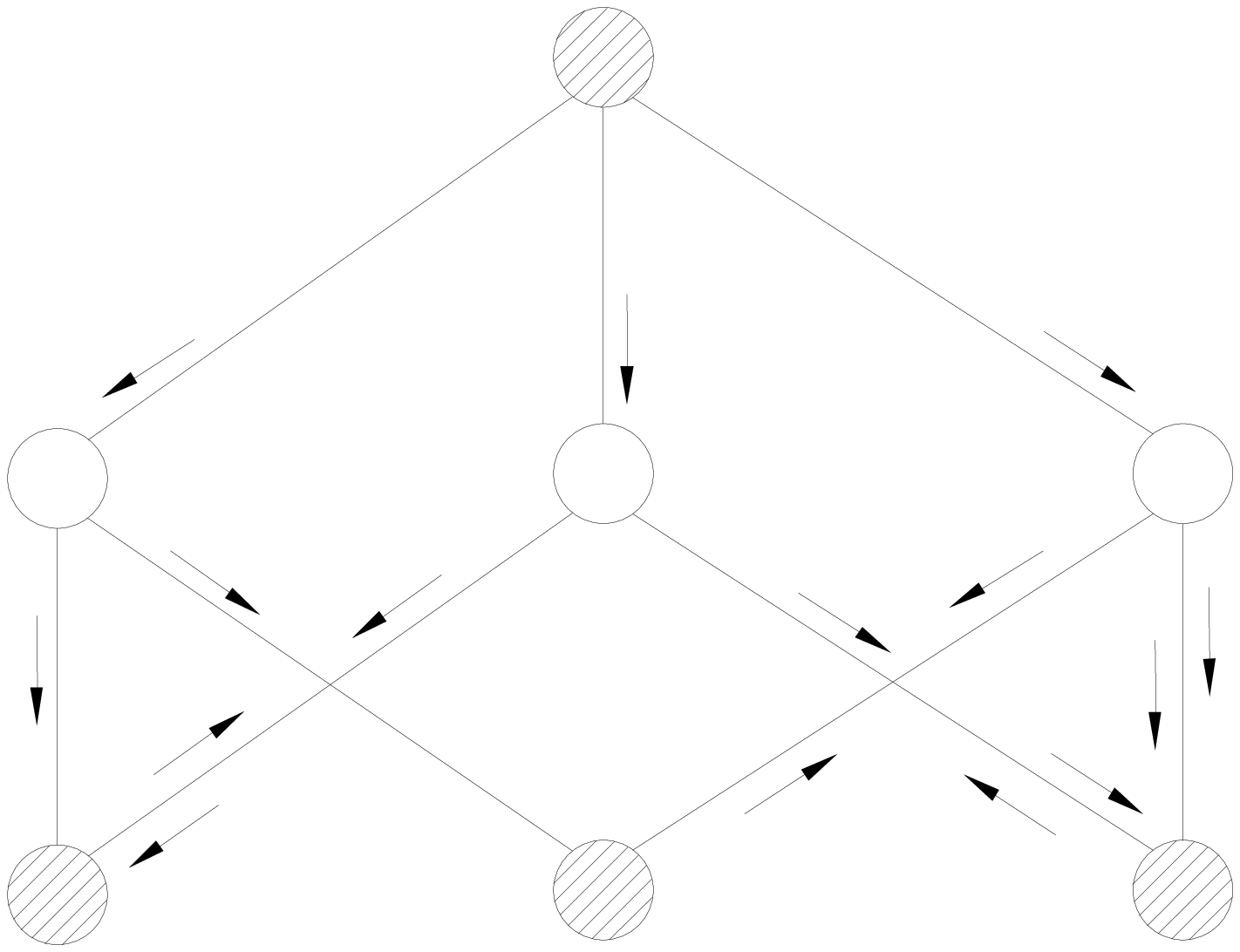} }
		\begin{small}
		\put(4.8,10){$\{ a_{0},a_{1},\ldots, a_{8} \}$}
		\put(16.3,10){$\{ a_{0},a_{1},\ldots, a_{8} \}$}
		\put(-0.1,7.5){$a_{0}$...$a_{4}$} \put(2.35,7.9){$a_{1}a_{2} a_{6}a_{7}a_{8}$} \put(7,7.5){$a_{4}a_{5} a_{6}a_{7}a_{8}$}
		\put(-2,4.9){$a_{0}$...$a_{4}$}  \put(0,5.4){$a_{0}$...$a_{4}$} 
		\put(2.6,4.5){$a_{6}a_{7}a_{8}$}    \put(4.8,5){$a_{1} a_{2} a_{8}$}
		\put(7.2,5.5){$a_{5}$...$a_{8}$}  \put(9,4.8){$a_{4}$...$a_{7}$}
		
		\put(8,3.5){$a_{0}$} \put(-0.5,3){$a_{0}$}  
		\put(8.4,2.3){$a_{5}$} \put(1,3){$a_{5}$} 
		\put(5,2.5){$a_{3}$} \put(8.6,4.2){$a_{3}$} 
		\put(8.5,0){ \includegraphics[width = 5.7cm,height = 4.54cm]{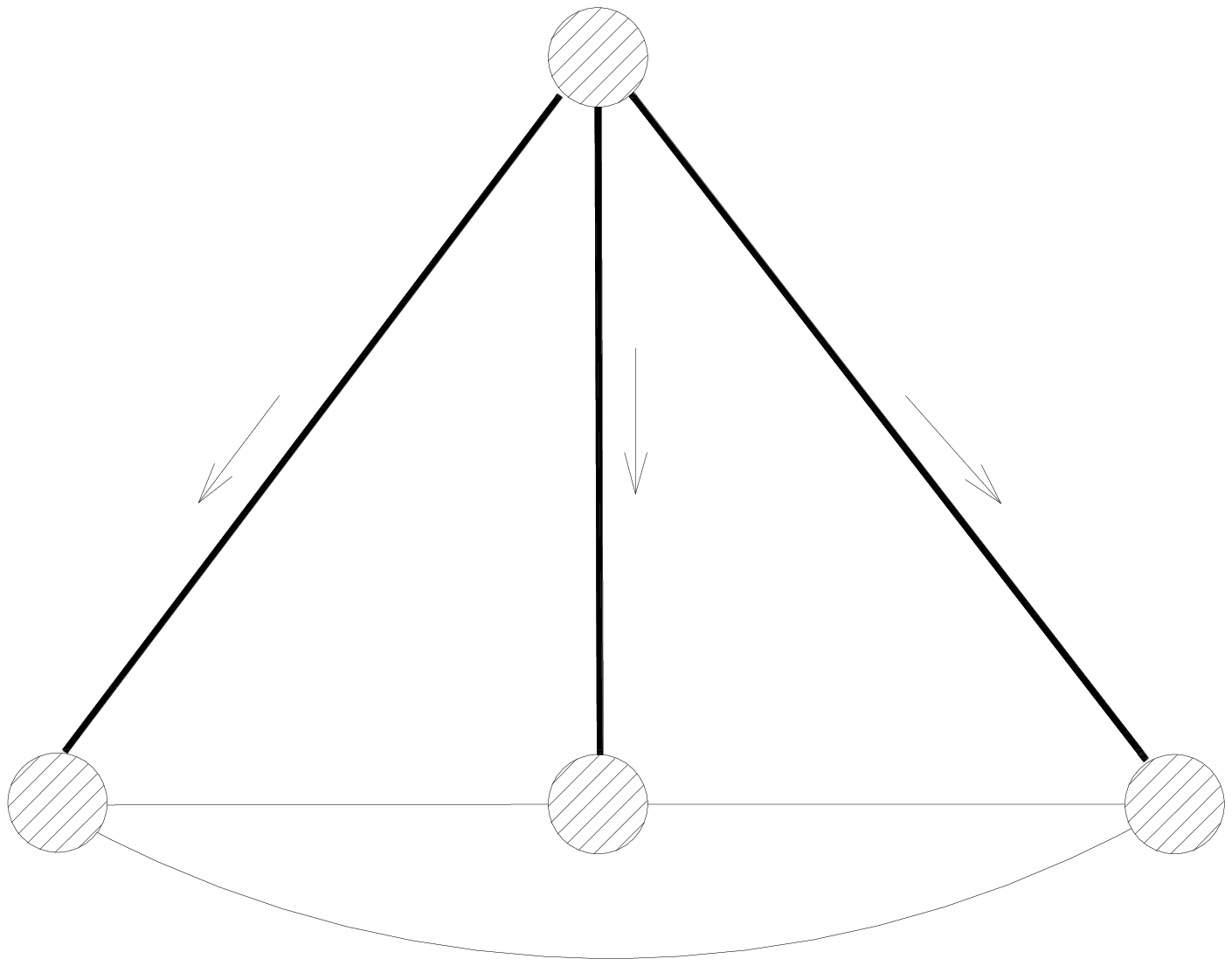} }
		\put(12.7,7){$a_{0}$...$a_{5}$} \put(15.1,7.5){$a_{2}$...$a_{7}$} \put(18.2,7){$a_{3}$...$a_{8}$}
		\put(12,3.3){$a_{0}a_{1} \rightarrow$} \put(16.5,3.2){$a_{0}a_{1} a_{2} \rightarrow$}
		\put(12.7,2.5){$a_{8} \rightarrow$} \put(14.5,1.8){$\leftarrow a_{6}a_{7} a_{8}$}
		\put(4,0){\mbox{(a)}} \put(16,0){\mbox{(b)}}
		\end{small}		
		\end{picture}
		\caption{Example 1: (a)-Multicast Network $G$. Also shown a fractional routing scheme that achieves a throughput of $1.8$. The source has $h=9$ messages $\{a_{0}, \ldots, a_{8}\}$ and the capacity of each link is upscaled by $n=5$. The symbols on the edges represent the messages carried by the edges and the arrows indicate the direction of the flow. All sinks will recover all the messages and thus a throughput of $\frac{9}{5} = 1.8$ is possible. , (b) Splitted graph $G^{\prime}$  together with a fractional routing scheme that achieves a throughput of $\frac{9}{6} = 1.5$.}
	\label{fig:Complete_Splitting}
\end{figure}

\textit{Example 1}: For the multicast network in Fig.\ref{fig:Complete_Splitting}(a), $\lambda(A) = 2$. The graph resulting from upscaling edge capacities by 2 and then performing suitable complete splitting at all relay (non-hatched) nodes is shown in Fig.\ref{fig:Complete_Splitting}(b), where thick edges have double the capacity of normal ones. $G^{\prime}$ is a broadcast network, and from theorem 3 with $A = V(G^{\prime})$ the routing capacity of $G^{\prime}$ can be bounded as $\pi_{f} \geq \frac{4}{3}$. From theorem 3, this also serves as a lower bound on the routing capacity of $G$ and thus the coding gain in $G$ is bounded by ${\mathfrak G}_{f} \leq 1.5$. Note that $\frac{4}{3}$ is a lower bound on the capacity, a routing scheme achieving a throughput of 1.5 for $G^{\prime}$ using fractional routing with $h=9$ and $n=6$ is shown in Fig.\ref{fig:Complete_Splitting}(b). For the network on $G$, a fractional routing throughput of $1.8$ is possible with $h=9$ and $n = 5$, Fig.\ref{fig:Complete_Splitting}(a).

\begin{figure}[htbp]
		\centering \setlength{\unitlength}{0.38cm}
		\begin{picture}(18,13)(0,-2)
		\begin{small}
		\put(3.2,9.8){$v_{0}$} \put(15,10.8){$v_{0}$} \put(14.8,4.3){$v_{0}$}
		\put(-4,-1){\includegraphics[width = 5.7cm,height = 4.56cm]{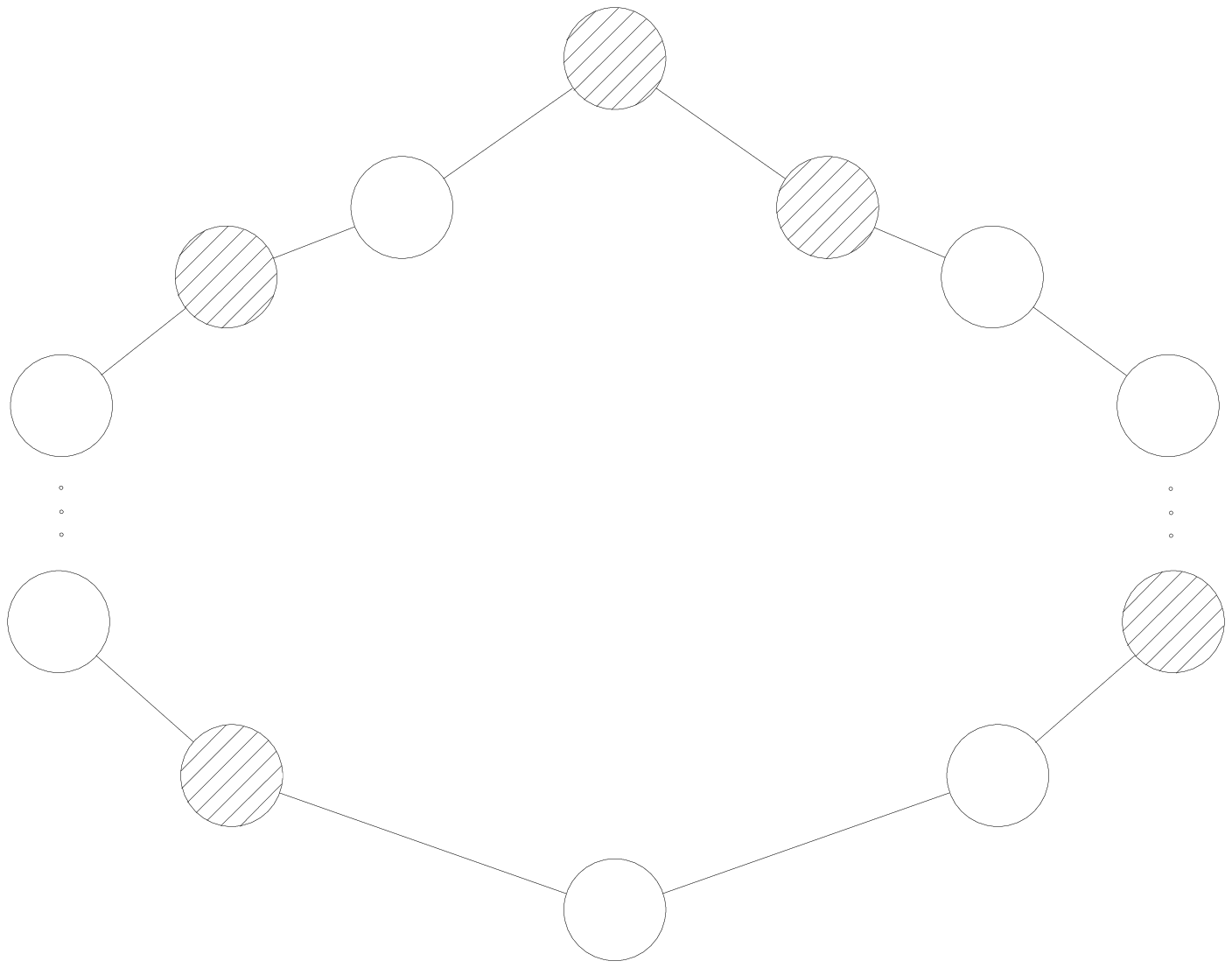} }
		\put(1.2,8.2){$x_{1}$} \put(-.2,7.8){$v_{1}$} \put(-1.8,6.5){$x_{2}$} \put(-.2,3.2){$v_{j}$}
		\put(4.8,8.3){$v_{|A|-1}$}
		\put(7.5,5){ \includegraphics[width = 5.7cm,height = 4.56cm]{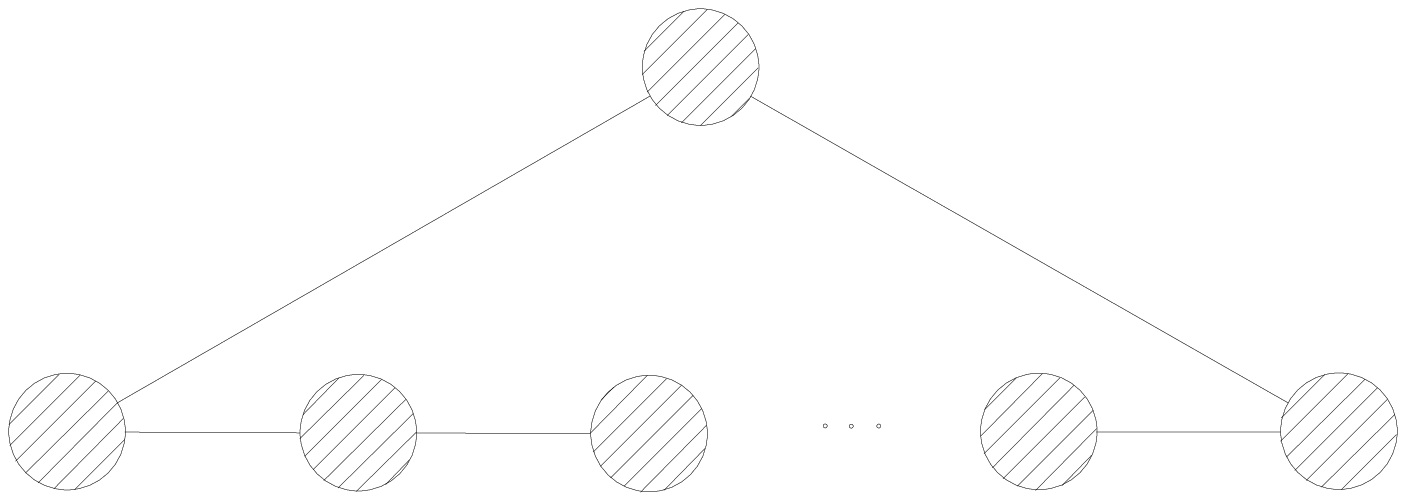} }
		\put(10,7.7){$v_{1}$} \put(12.2,7.7){$v_{2}$} \put(14.4,7.7){$v_{3}$} \put(16.8,7.7){$v_{|A|-2}$} 	     \put(19.2,6.1){$v_{|A|-1}$}
		\put(3,-1){\mbox{(a)}} \put(14.8,5.5){\mbox{(b)}} \put(14.8,-1.5){\mbox{(c)}}			
		\put(7.5,-1.5){\includegraphics[width = 5.7cm,height = 4.56cm]{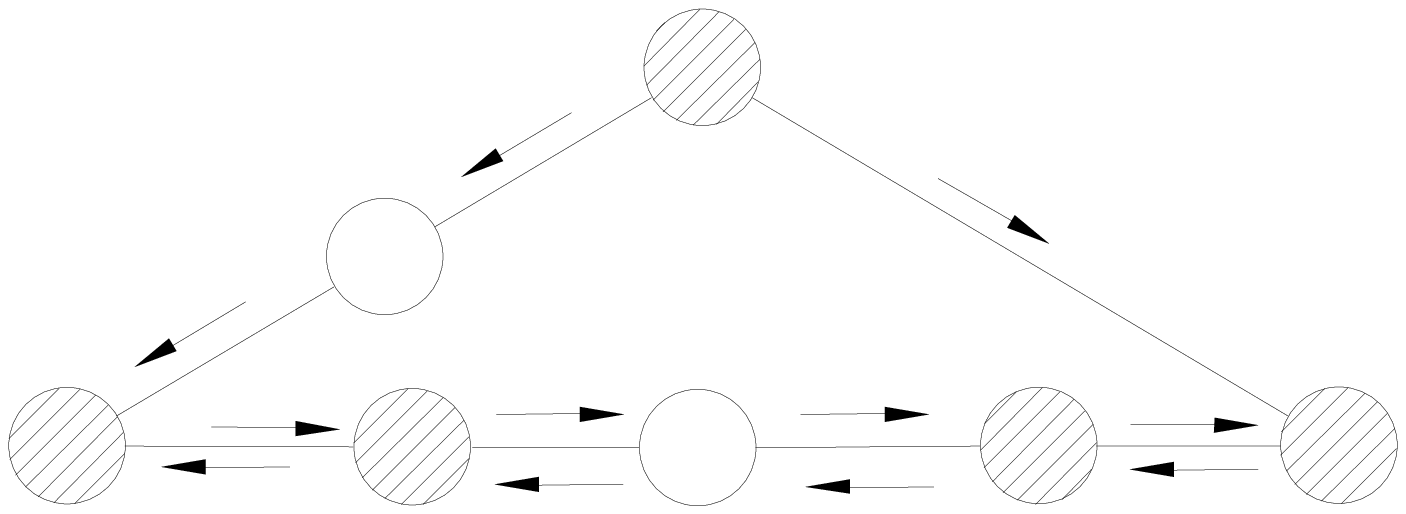} }
		\begin{tiny}
		\put(11.4,3.5){$a_{0}a_{1}a_{2}a_{3}$} \put(9,1.8){$a_{0}a_{1}a_{2}a_{3}$}
		\put(11,1){$a_{0}a{1}a_{2}$} \put(11,-.2){$a_{4}$}
		\put(13.3,1){$a_{0}a{1}$} \put(13.2,-.25){$a_{3}a_{4}$}
		\put(15.8,1){$a_{0}a{1}$} \put(15.5,-.25){$a_{3}a_{4}$}
		\put(18.2,.9){$a_{0}$} \put(18,-.25){$a_{2}a_{3}a_{4}$} 
		\put(16,3){$a_{1}a_{2}a_{3}a_{4}$}
		\end{tiny}
		\put(11.9,2.6){$x_{1}$} \put(9.6,-.5){$v_{1}$} \put(12.2,-.5){$v_{2}$} \put(14.8,1.1){$x_{2}$}           \put(17.3,1.1){$v_{3}$} \put(19.6,1.2){$v_{4}$}
		\end{small}	
		\end{picture}
		\caption{Example 2: (a)-Multicast Network $G$, (b) Splitted graph, $G^{\prime}$, (c) An instance of $G$ with 5 terminals and 2 relay nodes.}
	\label{fig:Complete_Splitting3}
\end{figure}

\textit{Example 2}: Fig.\ref{fig:Complete_Splitting3}(a) represents a multicast network with a set of terminals $A=\{v_{0}, v_{1}, \ldots, v_{|A|-1}\}$ and a set of relay nodes $X = \{x_{1}, \ldots, x_{|X|}\}$ where $X = V-A$ and $v_{0}$ is assumed to be the source node. The graph $G^{\prime}$ resulting from performing suitable complete splitting at all relay nodes is shown in Fig.\ref{fig:Complete_Splitting3}(b). From theorem 3 with $\lambda(A) = 2$, the routing capacity can be lower bounded as $\pi_{f} \geq \left(\frac{|A|}{|A|-1}\right)$ for both $G$ and $G^{\prime}$.

Next we show that for the network $G$ in Fig.\ref{fig:Complete_Splitting3}(a), $\gamma = \pi_{f} = \left(\frac{|A|}{|A|-1}\right)$. From \cite{Li2004:1}, $\gamma \leq \eta_{G}(A)$ where $\eta_{G}(A)$ is the edge \textit{strength} defined as
\begin{eqnarray}
\eta_{G}(A) = \mbox{min} \frac{E_{G}({\cal P})}{|{\cal P}| - 1} \nonumber
\end{eqnarray}
and the minimization is over all possible partitions ${\cal P} = \{V_{0}, V_{1}, \ldots, V_{|{\cal P}|-1}\}$ of $V(G)$ such that each \textit{component}, $V_{i}$, of the partition $\cal P$ contains at least one terminal, i.e. $V_{i} \cap A \neq \emptyset$. $E_{G}({\cal P})$ is the total 
capacity of edges between distinct components.
Let us choose a partition $\cal P$ such that $V_{i} \cap A = v_{i}$ $\forall i \in \{0,1,\ldots, |A|-1\}$. For such a partition, $|{\cal P}| = |A|$. Also since $G$ is a cycle with each edge having unit capacity, then $E_{G}({\cal P}) = |A|$. Because $\eta_{G}(A)$ is the minimum of the ratio $\frac{E_{G}({\cal P})}{|{\cal P}| - 1}$ over all possible partitions $\cal P$, then $\gamma \leq \eta_{G}(A) \leq \frac{|A|}{|A|-1}$. Combining this with the lower bound on $\pi_{f}$ obtained earlier from theorem 3 and the fact that $\pi_{f} \leq \gamma$, we obtain
\begin{eqnarray}
\frac{|A|}{|A|-1} \leq \pi_{f} \leq \gamma \leq \frac{|A|}{|A|-1} \nonumber
\end{eqnarray}    
Thus, for the multicast network $G$ shown in Fig.\ref{fig:Complete_Splitting3}(a), $\pi_{f} = \gamma  = \frac{|A|}{|A|-1}$.
 
A routing scheme achieving the fractional capacity can be advised as follows: Let $h = |A|$ and $n = |A|-1$. Node $v_{i}$ forwards the $|A|-(i+1)$ symbols $a_{0},a_{1},\ldots, a_{|A|-1-(i+1)}$ to terminal $v_{i+1}$, $\forall i \in \{0,1,\ldots,|A|-2\}$. Terminal $v_{0}$ sends $|A|-1$ symbols, $a_{1},\ldots,a_{|A|-1}$, to $v_{|A|-1}$. Finally, terminal $v_{i+1}$ forwards $i$ symbols, $a_{|A|-i},\ldots, a_{|A|-1}$, toward terminal $v_{i}$, $\forall i \in \{1,\ldots,|A|-2\}$. On the other hand, all relay nodes in $X$ do nothing but forwarding whatever they receive at their input edge to their output edge, Fig.\ref{fig:Complete_Splitting3}(c). Note that with such a routing scheme, each edge carries $|A|-1$ symbols and each terminal receives $|A|$ symbols. Achieving a routing throughput of $|A|/(|A|-1)$. Fig.\ref{fig:Complete_Splitting3}(c), shows an example with 5 terminals, $\{v_{0},v_{1},v_{2},v_{3},v_{4}\}$, and 2 relay nodes, $\{x_{1},x_{2}\}$.
\appendix
We use the notation $(a)_{b}$ to denote $a$ modulo $b$
\subsection{ $\Delta \in \{1,3,5,7\}$ for the multicast network with three terminals.}

$\!\!\!\!\!\!\!\!$ \textbf{Proof}: 
From (\ref{eq:any_int}) we can write
$
\lambda(A) = 4q+c, \hspace{.2cm} c \in \{0,1,2,3\}.
$
Also from (\ref{eq:Delta_3ter})
\begin{eqnarray}
\frac{6\lambda(A)-3}{8} = \left\lfloor \frac{6\lambda(A)-3}{8} \right\rfloor + \frac{\Delta}{8} \nonumber 
\end{eqnarray}
In other words,
\begin{eqnarray}
\Delta = (6\lambda(A)-3)_{8} = (3 \times 8q+6c-3)_{8} = (6c - 3)_{8} \nonumber
\end{eqnarray}
thus,
\begin{eqnarray}
\Delta =  \left \{ \begin{array}{cc}      5, & c=0 \\ 3, & c=1 \\ 1, & c=2 \\ 7, & c=3 
\end{array} \right. \nonumber
\end{eqnarray}
Note that for the case $\delta=1$, then $c=2$ and thus, $\Delta = 1$ as it was shown before.
\subsection{ For the multicast network with any number of terminals, $\Delta^{\prime} = 0$ if and only if $\Delta = 0$ and $\delta = 0$.}
$\!\!\!\!\!\!\!\!$ \textbf{Proof}: 
From (\ref{eq:lambda_general1})
\begin{eqnarray}
\lambda(A) = \frac{2k(|A|-1)+(|A|-2)}{|A|} - \frac{\Delta}{|A|} + \delta \nonumber
\end{eqnarray}
From which we can write,
\begin{eqnarray}
k \!\!\!\! &=&\!\!\!\! \frac{|A|\lambda(A)-|A|+2}{2(|A|-1)} + \frac{\Delta - |A| \delta}{2(|A|-1)} \nonumber \\
 &=&\!\!\!\! \left\lfloor \frac{|A|\lambda(A)-|A|+2}{2(|A|-1)} \right\rfloor + \frac{\Delta^{\prime} + \Delta - |A| \delta}{2(|A|-1)} \nonumber
\end{eqnarray}
where $\delta \in \{0,1\}$, $\Delta \in \{0,1,\ldots,|A|-1 \}$ and $\Delta^{\prime} \in \{0,1,\ldots,2(|A|-1)-1 \}$.
This shows that $\Delta^{\prime} + \Delta - |A| \delta$ is divisible by $2(|A|-1)$. Thus,
\begin{eqnarray}
\left( \Delta^{\prime} + \Delta \right)_{2(|A|-1)} = |A| \delta \nonumber
\end{eqnarray}
From this we note that if $\Delta = 0$ and $\delta = 0$, then $\Delta^{\prime} = 0$, which proves the 'if' part of the claim. Conversely, if $\Delta^{\prime} = 0$, then $\Delta = |A| \delta$, which shows that if  $\delta = 1$, then $\Delta = |A|$, a contradiction (since $\Delta \in \{0,1, \ldots, |A|-1\}$). Thus, $\delta = \Delta = 0$ and the claim follows. 
\bibliography{C:/Research_Latex/bibliographys/Network_Coding}



\end{document}